\documentclass[epjCONF,columns]{svjour}
\usepackage{graphics}
\usepackage[varg]{txfonts} 
\usepackage[latin1]{inputenc}
\usepackage{units}
\session-title{Hadron Collider Physics symposium (HCP-2011)}
\def\ZTau{\ensuremath{Z \rightarrow \tau\tau}}
\def\ET{\ensuremath{E_{\mathrm{T}}}}
\def\pt{\ensuremath{p_{\mathrm{T}}}}

\begin{document}
\title{Triggering On Hadronic Tau Decays: A challenge met by ATLAS}
\author{Marcus M. Morgenstern On behalf of the ATLAS Collaboration}
\institute{Technical University Dresden, Dresden, Germany}
\abstract{
The ATLAS experiment at the Large Hadron Collider (LHC) has been able
to collect $\unit[5.25]{fb^{-1}}$ of data in 2011. For many physics analyses both in context of the Standard
Model (SM) and Beyond the Standard Model (BSM) theories such as Higgs boson
searches, tau leptons play an important role. Thus, triggering on
hadronic tau decays is an essential ingredient for the success of
those measurements. This contribution will summarize the developed efforts to
meet this challenge. Efficiency measurements using data taken in 2011
at a center-of-mass energy of \unit[7]{TeV}
are described and results are presented. An outlook on further
developments of the tau trigger algorithms, to match future
requirements and higher instantaneous luminosities are summarised in the end.
} 
\maketitle
\section{Introduction}
\label{intro}
Tau leptons play a crucial role in many physics processes investigated by
the ATLAS experiment at the Large Hadron Collider (LHC), both within the Standard Model and in
models beyond the SM. In particular for Higgs boson decays in
the SM at low masses and in the Minimal Supersymmetric extension of
the SM (MSSM) the branching fraction to taus is expected to be in the order of 10\%. For this purpose it is important to
trigger on tau decays efficiently. The tau trigger is aimed at efficiently
rejecting background events with fake tau candidates while keeping the
acceptance for real taus as high as possible. Tau leptons decay hadronically
65\% of the time and leptonically 35\% of the time. Only hadronic tau decays are
considered in this context. Taus tend to decay into one or three
charged hadrons, mainly pions and to a small amount into kaons, plus
additional neutral pions or kaons. Thus,
the tau decay is characterised by one or three charged tracks and
narrow energy deposists in the electromagnetic and hadronic
calorimeter. These decay patterns become manifest in isolation criteria
applied to distinguish tau decays from background, mainly arising from
QCD multijet contributions.

\section{The ATLAS Tau Trigger}
\label{sec:1}
\begin{figure}[h!]
\resizebox{1.\columnwidth}{!}{\includegraphics{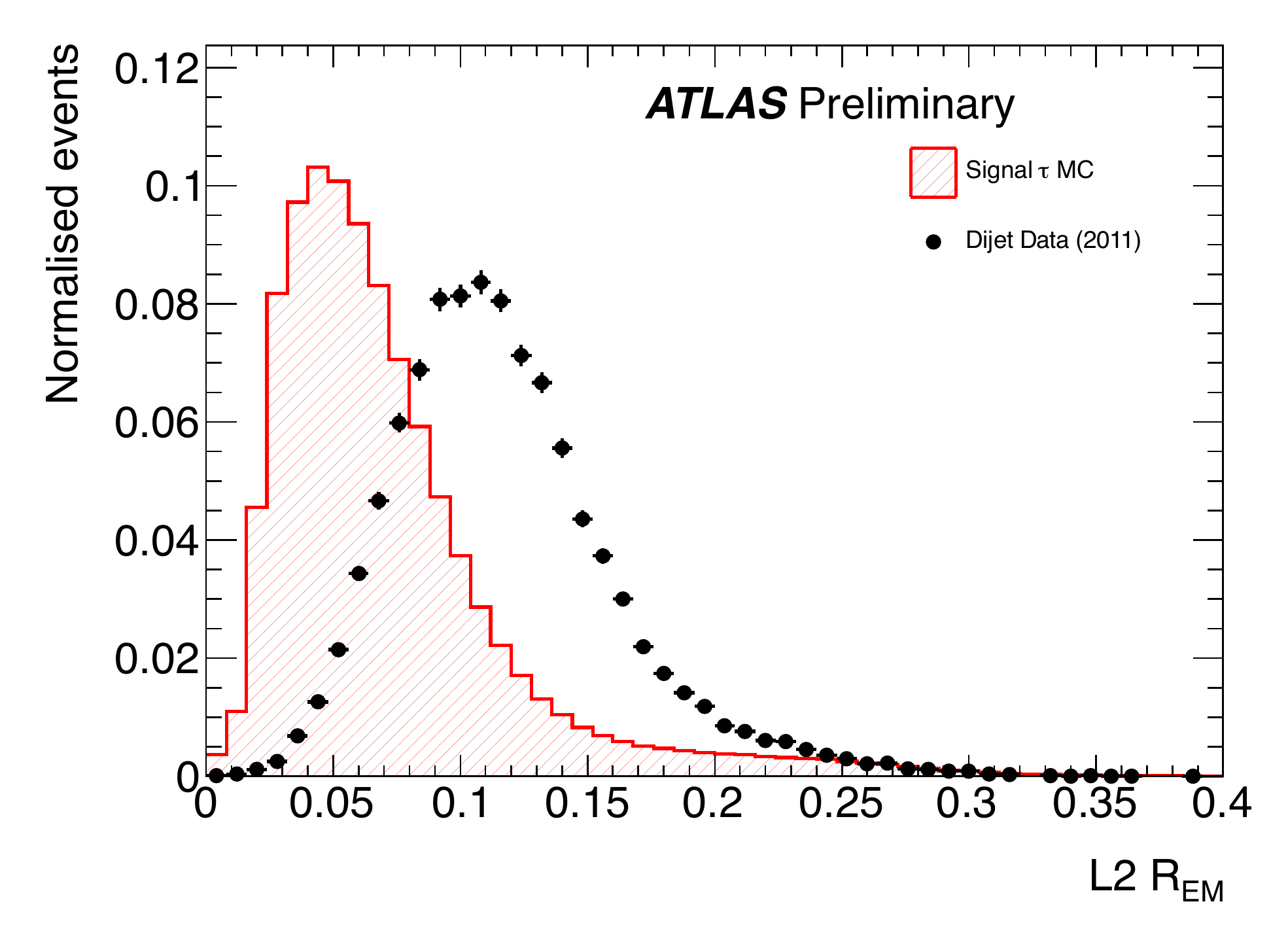}}
\caption{Electromagnetic radius, $R_{em}$, for trigger taus
in signal Monte Carlo and QCD di-jet data. The latter serves as
background estimate \cite{RefTrigPlots}.}
\label{fig:IDVar1}       
\end{figure}

\begin{figure}[h!]
\resizebox{1.\columnwidth}{!}{\includegraphics{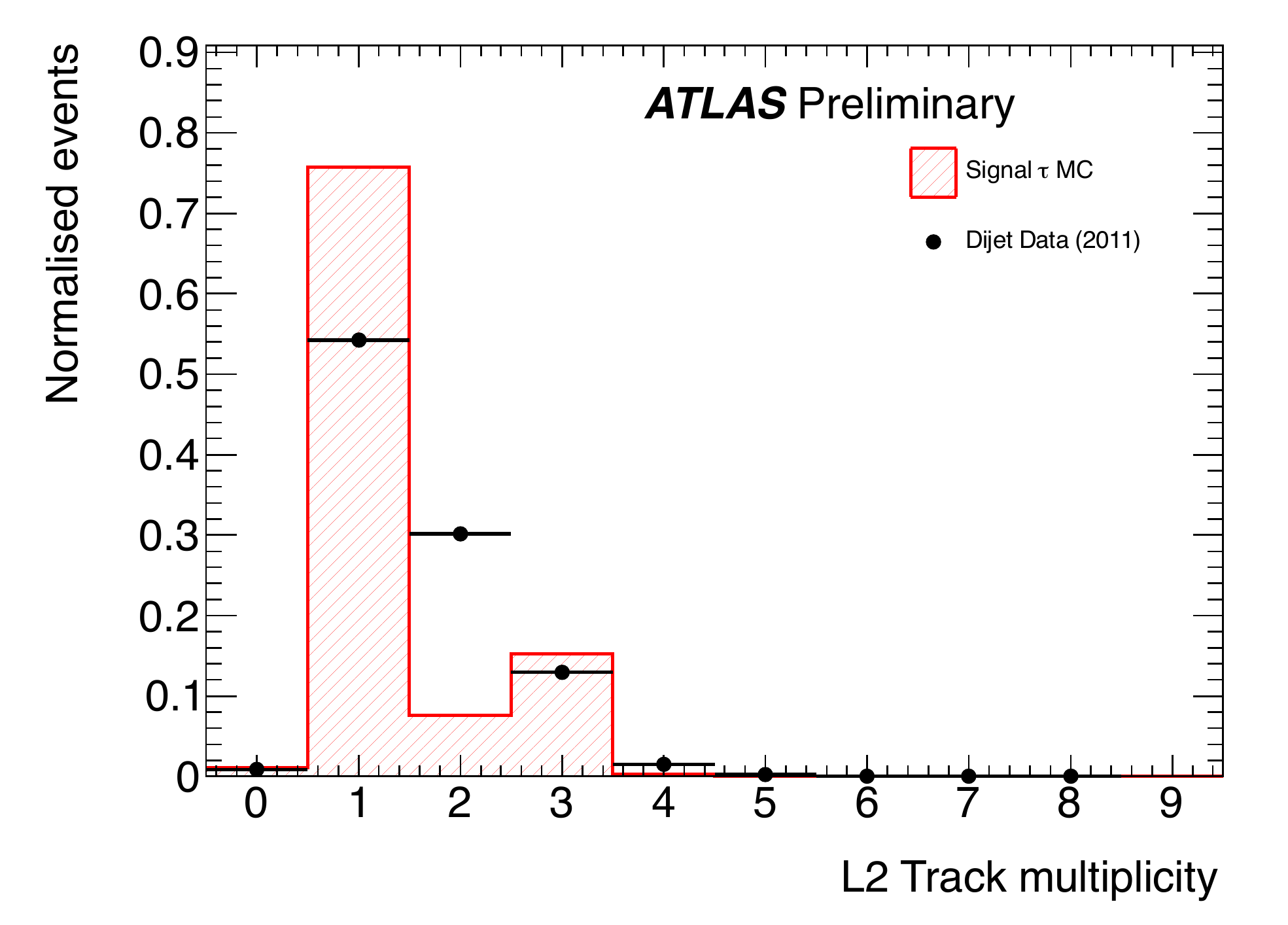}}
\caption{Number of tracks of L2 taus for trigger taus
in signal Monte Carlo and QCD di-jet data. The latter serves as
background estimate \cite{RefTrigPlots}.}
\label{fig:IDVar2}       
\end{figure}

The ATLAS trigger system \cite{RefTrigger} is aimed at reducing the initial collision rate of
\unit[40]{MHz} to a reliable data rate for permanent storage of around
\unit[200-400]{Hz}. This is achivied by a three-level system. The first
system is the level 1 (L1) trigger system. It is a hardware based
system which uses information from the electromagnetic (EM) and hadronic
calorimeter systems, provided as trigger towers of size $\Delta \eta \times
\Delta \phi = \unit[0.1]{} \times \unit[0.1]{}$. At level 1 a tau
lepton is identified by applying a given transverse energy, \ET{}, cut corresponding to
the given trigger item, e.g. for \emph{L1\_tau15} a \ET{} threshold
strictly above \unit[15]{GeV} is applied. The energy is reconstructed by the sum of
energy deposits in the highest $\unit[2]{} \times \unit[1]{}$ pair of EM towers and
the sum of energy deposits of $\unit[2]{} \times \unit[2]{}$ hadronic
towers behind the EM layers. In addition to
the \ET{} threshold a cut on the energy in the isolation region can be
applied which is measured as the sum of energy deposits in an
isolation region in a ring of $\unit[4]{} \times \unit[4]{}$ around the
core region of $\unit[2]{} \times \unit[2]{}$ towers. A so-called
Region of Interest (RoI) is built during the L1 processing and is
passed to the Higher Level Trigger (HLT), if the event is accepted. To
stay below the rate limit of \unit[75]{kHz} a hardware based pre-scale (PS) factor can be applied.
Next, the level 2 (L2) system performs track reconstruction within the
RoI in
addition to calorimeter information (using the full detector
granularity in the RoI) to identify tau leptons. Identification variables
measuring tau decay properties, such as the narrowness of the tau
decay products or the number of reconstructed tracks, are used to
provide a sufficient discrimination power against QCD jets. By setting requirements
on those identification variables, several working points
corresponding to given signal efficiencies are
defined. In addition a PS can be applied. Figure~\ref{fig:IDVar1} presents the electromagnetic radius,
$R_{\mathrm{EM}},$ which measures the lateral shape of the tau
decay. Figure~\ref{fig:IDVar2} shows the number of tracks for L2 taus obtained in
signal Monte Carlo (MC), and background QCD di-jet data. 

The last trigger step is accomplished by the event filter (EF), which
runs on fully built events.. Reconstruction and
identification algorithms similar to the ones used for offline
analysis are utilised \cite{RefTau}. Depending on the trigger chain a PS factor is
applied to match the final ouput rate of around \unit[200-400]{Hz}.

\section{Tau Trigger Efficiency}
\label{sec:2}

For physics analyses it is mandatory to know the trigger
efficiency. This is the probability for an offline-reconstructed and
identified tau to pass the trigger. For this purpose several studies measuring the
trigger efficiency have been performed using both Monte Carlo
simulated and real data events.

\begin{figure}[h!]
\resizebox{1.\columnwidth}{!}{\includegraphics{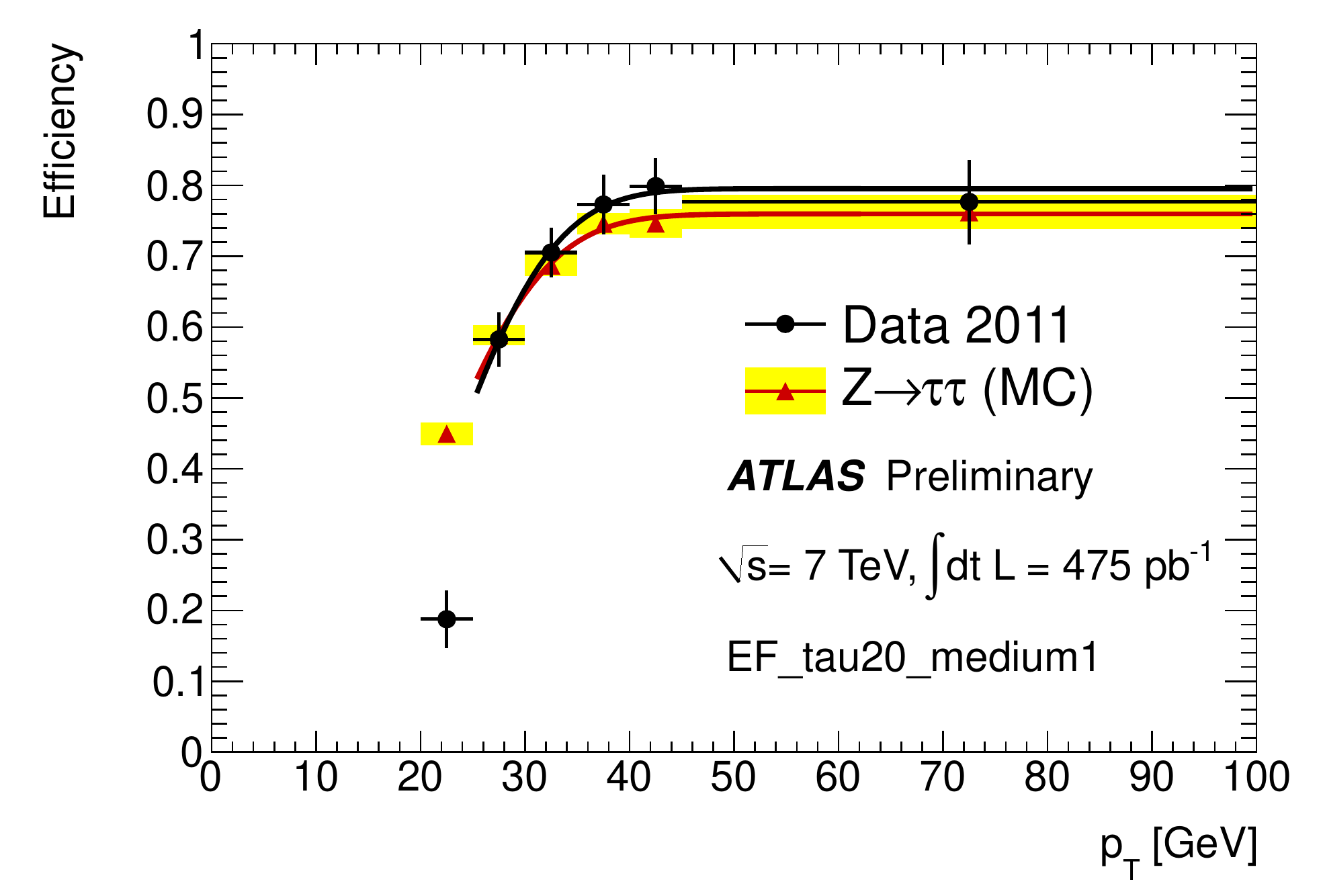}}
\caption{Tau trigger efficiency measured in \ZTau{} decays using 2011
  data by applying a tag-and-probe method for the EF \emph{tau\_20\_medium}
item. The efficieny is
measured w.r.t. offline tau candidates passing medium identification
requirements as a function of the offline tau \pt{} \cite{RefTrigPlots}.}
\label{fig:trigeff1}       
\end{figure}

\begin{figure}[h!]
\resizebox{1.\columnwidth}{!}{\includegraphics{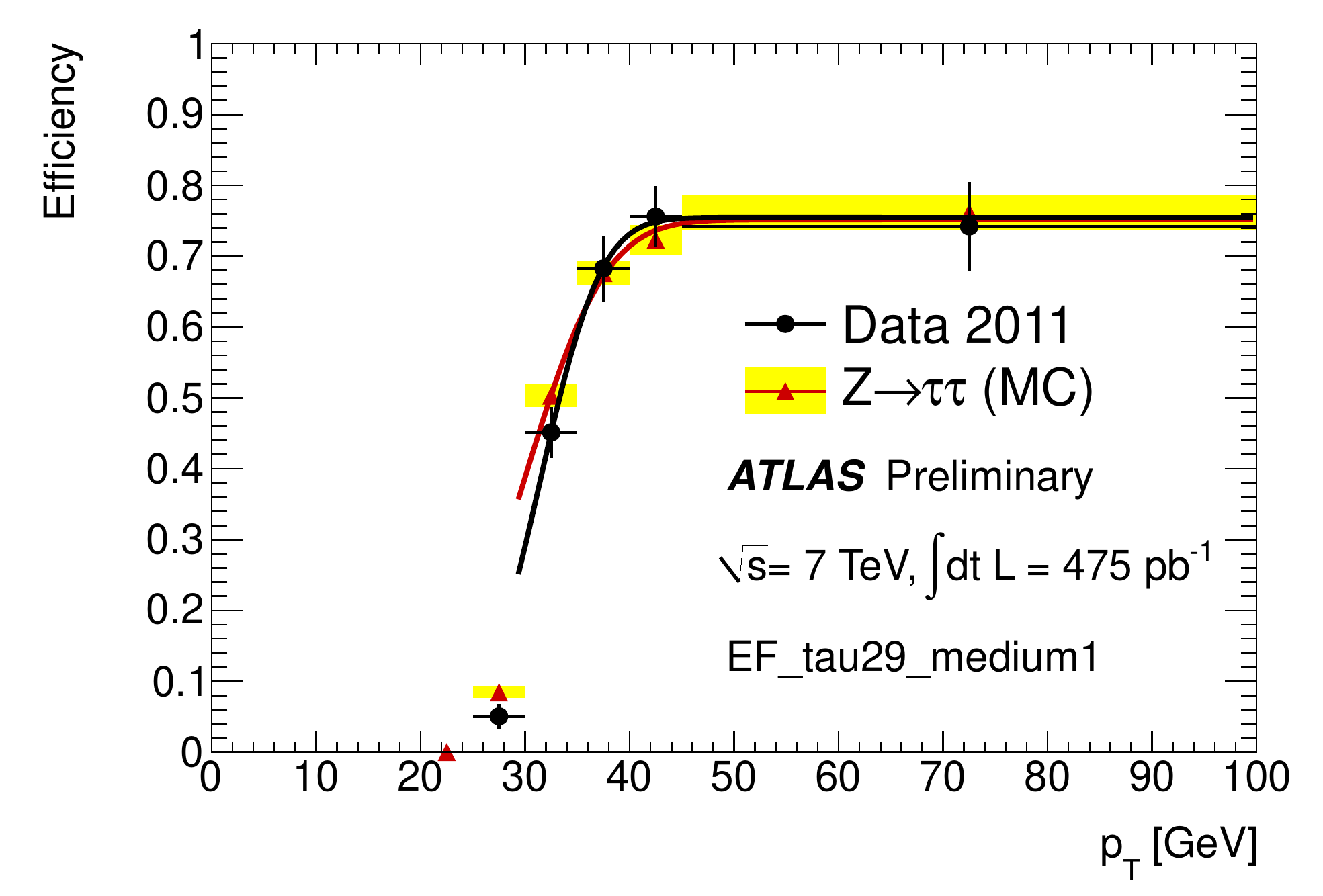}}
\caption{Tau trigger efficiency measured in \ZTau{} decays using 2011
  data by applying a tag-and-probe method for the EF \emph{tau\_29\_medium}. The efficieny is
measured w.r.t. offline tau candidates passing medium identification
requirements as a function of the offline tau \pt{} \cite{RefTrigPlots}.}
\label{fig:trigeff2}       
\end{figure}

To measure the tau trigger efficiency a study based on \ZTau{} MC
events has been performed. The efficiency is measured for several
trigger items as a function of the transverse energy, \ET{}. Since MC simulation will not necessarily give a reasonable description of
the new energy frontier of the LHC collisions, data driven measurements
of the trigger efficiency are essential. Hence, a study using $Z$ boson
decays in real data has been performed.  Herein the trigger efficiency is measured by a
tag-and-probe method using data of the 2011 run at a center of
mass energy of $\sqrt{s} = \unit[7]{TeV}$. An event is
tagged by either an electron or a muon, by applying an unbiased electron or
muon trigger, respectively. On the probe side tau leptons which are
reconstructed and identified by the offline algorithms are used to measure the tau trigger
efficiency. The amount of data analysed in
this measurement corresponds to roughly $\unit[475]{pb^{-1}}$. The results for the EF
\emph{tau\_20\_medium} and \\\emph{tau\_29\_medium} items are shown in
Figure~\ref{fig:trigeff1} and \ref{fig:trigeff2}. For these items medium tau trigger
identification is applied as well as a cut on \ET{} of $\unit[20]{GeV}$
and $\unit[29]{GeV}$, respectively.

To gain knowledge of the tau trigger efficiencies at very high
energies, where a SM sample of real tau leptons is limited in a few
$\unit{pb^{-1}}$ of data, QCD di-jet events can be used to achieve a
reasonable precision. Such a measurement is based on the idea that a
jet faking a reconstructed and identified offline tau lepton should
fire the tau trigger at high \pt{} at the same rate as real tau
leptons. Hence, a di-jet sample is selected in collision data and the trigger efficiency is measured by a tag-and-probe method.
\section{Future prospects for 2012 data}
\label{sec:2}

\begin{figure}
\resizebox{1.\columnwidth}{!}{\includegraphics{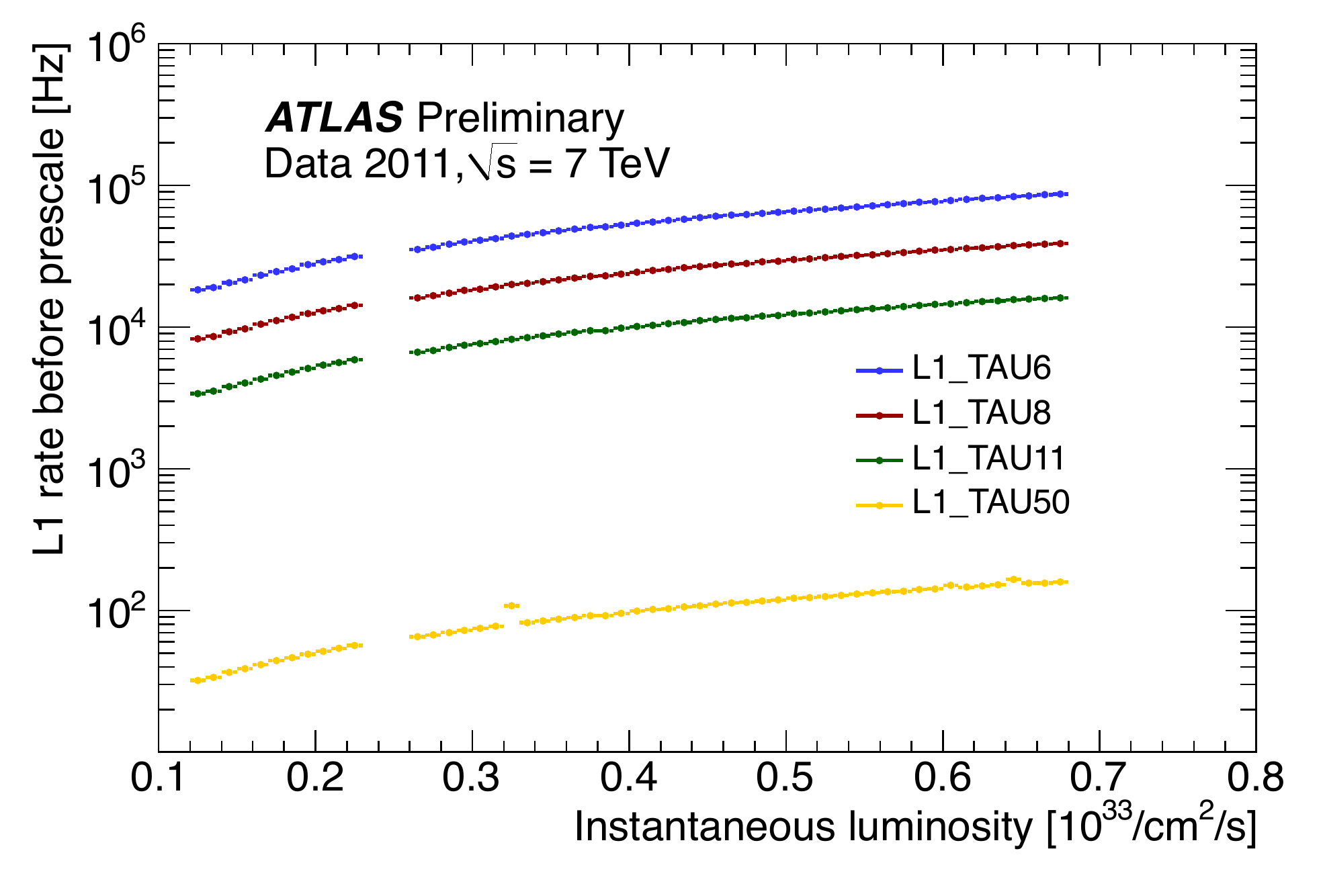}}
\caption{Evolution of the event rate as a function of the instantaneous
luminosity for L1 items as measured in 2011
data \cite{RefTrigPlots}.}
\label{fig:rate1}       
\end{figure}

\begin{figure}
\resizebox{1.\columnwidth}{!}{\includegraphics{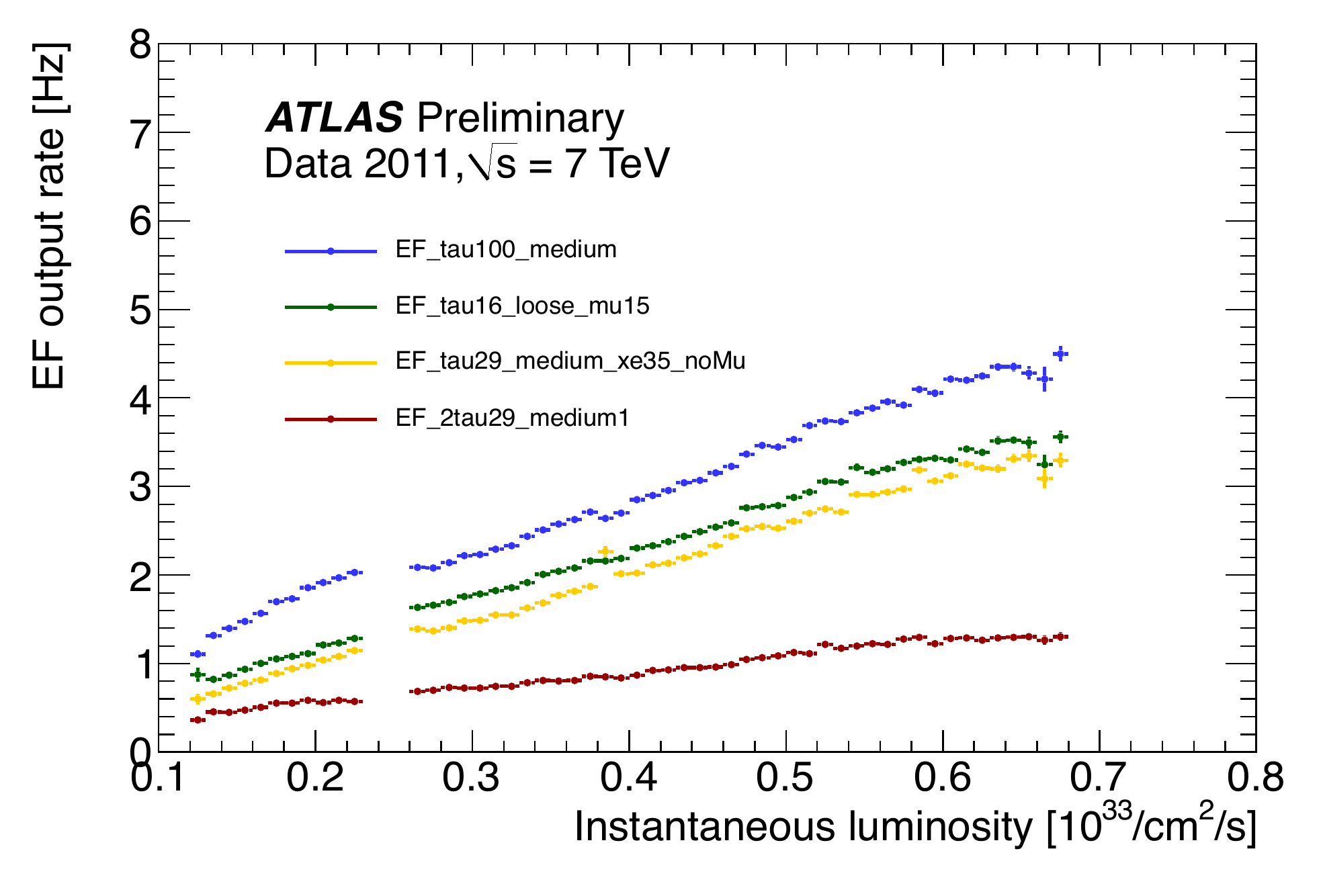}}
\caption{Evolution of the event rate as a function of the instantaneous
luminosity for EF items as measured in 2011
data \cite{RefTrigPlots}.}
\label{fig:rate2}       
\end{figure}

After a maintance stop in early 2012 the LHC will restart
proton-proton collisions in early April. To deal with the expected
increase in instantaneous
luminosity, and thus with a larger pile-up contribution as well as
higher trigger rates, improvements in
the selection algorithms of the tau trigger have to be
implemented. The evolution of the event rate as a function of the
instantaneous luminosity is shown in Figure~\ref{fig:rate1} and \ref{fig:rate2} for
different L1 and EF items, respectively. 

Several options are available to match the
given fixed output rate. The straightforward way is to increase the \ET{} of the items without
prescale. Furthermore, the isolation criteria can be tightened. Both
result in a lower acceptance for physics analysis. A different
approach to obey the necessary rejection rate is to apply more
advanced algorithms, as used in offline identification. Due to the higher rejection power at a given signal efficiency working
point, mulitvariate techniques could be a reasonable alternative to
the current cut based implementation of the tau trigger
algorithms. This is planned for the 2012 data taking.

\end{document}